\documentclass[aps,pre,amsmath,amssymb,twocolumn,groupedaddress,showpacs]{revtex4}

\usepackage{graphicx}
\newcommand{\be}{\begin{equation}}
\newcommand{\ee}{\end{equation}}
\newcommand{\ba}{\begin{array}{l}}
\newcommand{\ea}{\end{array}}
\newcommand{\re}[1]{(\ref{#1})}
\newcommand{\ci}[1]{\cite{#1}}
\newcommand{\banonum}{\begin{eqnarray*}}
\newcommand{\eanonum}{\end{eqnarray*}}
\newcommand{\baa}{\begin{eqnarray}}
\newcommand{\eaa}{\end{eqnarray}}
\newcommand{\bfr}{\begin{flushright}}
\newcommand{\efr}{\end{flushright}}
\newcommand{\bfl}{\begin{flushleft}}
\newcommand{\efl}{\end{flushleft}}
\newcommand{\lab}[1]{\label{#1}}
\renewcommand{\baselinestretch}{1.2}
\begin{document}
%\large

\title{High harmonic generation in time-dependent quantum box}
\author{S. Z. Rakhmanov$^a$, H. Abduvakhidov$^a$, O. V. Karpova$^b$ and   D. U. Matrasulov$^b$}
\affiliation{$^a$National University of Uzbekistan, Vuzgorodok,
Tashkent 100174,Uzbekistan\\
  $^b$ Turin Polytechnic University in Tashkent, 17
Niyazov Str., 100095,  Tashkent, Uzbekistan}

\begin{abstract}
We consider optical harmonic generation in time-dependent box
driven by external time-periodic field. Two types of the external
field is considered: Time-periodic optical field and a field
created by harmonically oscillating wall of the box. The latter is
treated on in terms of the Schrodinger equation with
time-dependent boundary conditions.
\end{abstract}
\maketitle

\section{Introduction}
Quantum particle dynamics in time-dependent traps such as hard
wall box, cavity and well attracted much attention during past few
decades \ci{TDT1}-\ci{TDT4}. Modern technological developments
allow trapping and manipulating of particles  in time-dependent
potentials, including those which can create very high barriers
making impossible particle escape.  Manipulation of the particle
dynamics  is of practical importance in such field as metrology
and quantum information processing. Possibility for creating of
time-dependent traps and confining there particles and atomic
bound states have been discussed recently in different contexts
\ci{TDT3,TDT4}. In most of the cases confinement  is created by
optical potential that causes additional effects of particle-wall
interaction, such as pressure, compressing of the confined
particle, acceleration, optical harmonic generation, etc. In other
words,interaction of confined particles with the moving boundary
causes modification of the boundary conditions making them
dynamical that leads to different dynamical effects which does not
appear in case of the static confinement.  The problem of moving
boundaries in quantum mechanics is treated in terms of the
Schr\"{o}dinger equation with time-dependent boundary conditions.
Earlier, the quantum dynamics of a particle confined in a
time-dependent box was studied in different contexts (see Ref.
\ci{doescher}-\ci{Our03}).

In this paper we consider optical harmonic generation caused by
the interaction of confined particles with the harmonically
oscillating wall by focusing in the role of confinement. We note
that earlier, the optical harmonic generation in confined systems
has been studied in different context (see, e.g.,
Refs.\ci{Ahn88}-\ci{Kotova} and book \ci{Boydbook}). Pioneering
treatment of the nonlinear effects, including harmonic generation
dates back to the Ref.\ci{Bois}, where optical nonlinearities in
asymmetric quantum wells due to resonant intersubband transitions
are studied within the compact density-matrix approach. In
\ci{Zaluz} similar problem is studied for nonparabolic two-level
quantum well systems by taking into account depolarization
effects. In \ci{Milanovic} a systematic procedure  for the optimal
design of quantum-well structures, which provide maximal resonant
second-order susceptibility is proposed. Different nonlinear
optical properties of in semiconductor quantum well are studied in
\ci{Liu,Zaluz1,Kotova}. Optical rectification, second- and
third-harmonic generations in a semispherical quantum dot placed
at the center of a cubic quantum box are studied in
\ci{Mohammadi}. Despite the fact that different aspect of
nonlinear optical phenomena in confined quantum systems are
studied, most of the researches are restricted  by considering the
case of static confinement. However, dynamical confinement appears
in many nanoscale systems and low-dimensional functional
materials, where optical processes play important role. Here we
study the role of dynamical confinement in high harmonic
generation in confined quantum system by considering static and
dynamic confinements. The latter is assumed to be created by
moving wall of the box. This can be achieved, e.g., when
confinement is caused by optical field, e.g., in atom optic
billiards or tweezers.  Since in most cases such oscillating trap
can be created by optical field, interaction of the moving wall
with the confined particles can cause nonlinear optical phenomena
such as harmonic generation. The paper is organized as follows. In
the next section we consider the problem of optical harmonic
generation in 1D quantum box driven by external monochromatic
field. In section III similar problem is considered in a quantum
box with harmonically oscillating walls. Section IV p[resents some
concluding remarks.

\section{Harmonic generation in driven quantum box}
Consider first the case of quantum particle confined in a
impenetrable box of size $L$ and driven by external linearly
polarized monochromatic field with the strength $F$ and frequency
$\omega_0$. Such system can be described by time-dependent
Schrodinger equation which is given by ($\hbar =m=e=1$)
\begin{equation}
i\frac{\partial \Psi}{\partial t}=(-\frac{1}{2}\frac{d^2}{dx^2} -
Fxcos\omega_0 t) \Psi, \;\; 0 < x < L, \label{Schr1}
\end{equation}
for which the (box) boundary conditions for $\Psi(x,t)$ are imposed
as \be \Psi(0,t) = \Psi(L,t) =0. \lab{bc001}\ee

Solutions of Eq.\re{Schr1} can be expanded in terms of the complete
set of the unperturbed box eigenfunctions as \be \Psi(x,t) =\sum_n
C_n(t)u_n(x),\label{exp01} \ee where $u_n(x)$ are he eigenfunctions
of the quantum particle confined in 1D box to be found from
\begin{equation}
H_0 u_n=E_n u_n. \label{tisch}
\end{equation}

Explicitly, $u_n$ and $E_n$ can be written as
\begin{equation}
u_n=\sqrt{\frac{2}{L}}\sin \frac{\pi n x}{L}, \label{wfun}
\end{equation}
and
\begin{equation}
E_n=\frac{\pi^2 n^2}{L^2}.  \label{ener}
\end{equation}
The functions $u_n$ fulfill the orthonormality condition given by
\begin{equation}
\int {u^{\ast}_m u_n dx}=\delta_{mn},  \label{normcond}
\end{equation}

For expansion coefficients, $C_n(t)$ using Eq.\re{Schr1} we have
system of first order differential equations given by

\be i\dot C_n(t) = E_nC_n - F\sum_n C_m(t)V_{mn}\cos\omega_0 t
\lab{exp02},\ee where \be V_{mn} = \int xu_m^*u_ndx.\lab{matrix}
\ee We are interested in the study of optical harmonic generation
in the system described by Eq.\re{Schr1}. The main physical
characteristics of such process is the average dipole moment which
is given by
$$
\bar d(t) = -<\Psi(x,t)|ex|\Psi(x,t)>,
$$
where $e$ is the electron charge.

\begin{figure}[t!]
\includegraphics[totalheight=0.23\textheight]{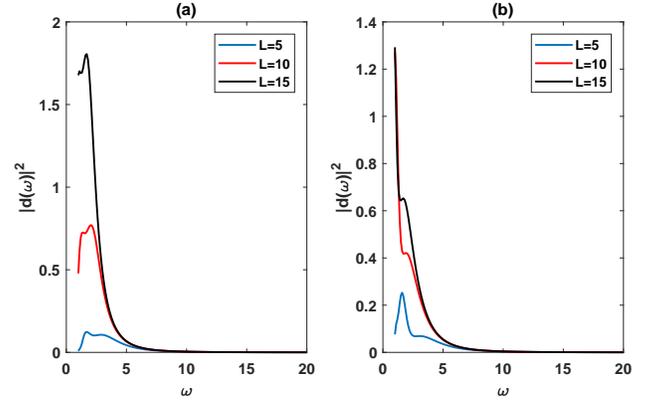}
 \caption{ (Color online) Harmonic generation spectrum for a quantum box driven by external, linearly polarized monochromatic potential at $\omega_0=0.1 $ (a) and $\omega_0=1 $ (b)} \label{fig:1}
\end{figure}

\begin{figure}[t!]
\includegraphics[totalheight=0.23\textheight]{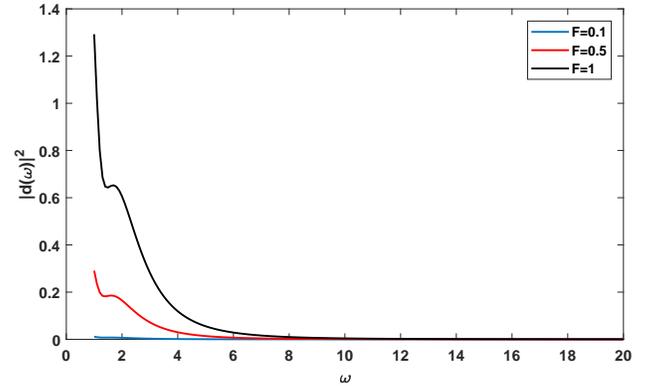}
 \caption{ (Color online) Harmonic generation spectrum for a quantum box driven by external, linearly polarized monochromatic potential for different external field amplitudes at $\omega_0=1 $ and $L=15 $ } \label{fig:2}
\end{figure}

Using Eqs.\re{exp01} and \re{normcond} one can write the average
dipole moment as
$$
\bar d(t) =-\sum_{m,n} C_m^*C_n(t) V_{mn}, \lab{dip02}
$$
where $V_{mn}$ is given by Eq.\re{matrix}. The spectrum of harmonic
generation is given by \be \bar d(\omega) =\frac{1}{T}\int_{0}^{T}
e^{-i\omega t}\bar d(t)dt, \lab{spectr1}\ee where $T=2\pi/\omega$.

In Fig.1 $|d(\omega)|^2$ which determines the intensity of
harmonic generation, is plotted at different values of the box
size, $L$ for the values of external field amplitude and
frequency,$F= 1$ and $\omega_0=0.1 $ (a), $\omega_0=1 $ (b). The
intensity decreases, as the harmonic order increases for all
values of $L$. Also, increasing of the  frequency of external
field does not change the situation, i.e., $|d(\omega)|^2$.  In
Fig.2 $|d(\omega)|^2$ is plotted for the different values of
external field strength, $F$ for $L= 15$ and $\omega_0=1 $. The
intensity decreases as the harmonic order increases. However, for
higher field strengths the decrease is slower compared to lower
values of $F$. All this implies that driven static quantum box is
not interesting from the viewpoint of higher order harmonic and
attosecond pulse generation. Therefore in  section III we consider
harmonic generation quantum box  induced by dynamical confinement.

\section{Time-dependent quantum box}

Consider quantum article confined in a 1D box with moving (right)
wall, i.e., the position of the right wall is given by $L(t)$.
Dynamics of such particle is governed by the time-dependent
Schrodinger equation given by
\begin{equation}\large
i \hbar \frac {\partial \psi}{\partial t}=-\frac{1}{2}
\frac{\partial^2 \psi}{\partial x^2} \label{shred01}
\end{equation}
The boundary conditions for this equation are imposed as
\begin{equation}\large
\psi (x,t)|_{x=0}=\psi (x,t)|_{x=L(t)}=0
 \label{bc}
\end{equation}
To solve Eq.\re{shred01} one should transform the boundary
conditions into time-independent (static) form. This can be done
by  new coordinate, $y$ which is given by
$$
y =\frac{x}{L(t)},
$$
and using the transformation of the wave function
\begin{equation}\large
\psi(y,t)=\sqrt {\frac{2}{L}} \,e^{\frac {i L\dot L
y^2}{2}}\varphi (y,t). \label{wavfun}
\end{equation}
together with the time scaling given by
$$
\tau =\int_0^t\frac{ds}{[L(s)]^2},
$$
we can rewrite  Eq.\re{shred01} as
\begin{equation}\large
i \frac {\partial \varphi}{\partial \tau}=-\frac{1}{2}
\frac{\partial^2 \varphi}{\partial y^2}+\frac{1}{2}L^3 \ddot
{L}y^2\varphi
 \label{shred03}
\end{equation}

Time and coordinate variables in Eq.\re{shred03} is possible only
for the case, when $L(t)$ fulfills
$$
L^3 \ddot{L} =const.
$$
For arbitrary time-dependence of $L(t)$  the problem can be solved
only numerically, e.g., by expanding $\varphi(y,t)$ in terms of
the complete set of eigenfunctions of a quantum box of unit size:
\begin{equation}\large
\varphi (y,t)=\sum_n C_n(t) \phi_n (y), \label{func}
\end{equation}
where $\phi_n (y)$ can be found from the following Schrodinger
equation:
\begin{equation}\large
-\frac{1}{2} \frac {d^2\phi _n}{dy^2} =E_n \phi _n,
\label{newcoor}
\end{equation}
with the boundary conditions
\begin{equation}\large
\phi _n|_{y=0}=\phi_n |_{y=1}=0.
 \label{newbc}
\end{equation}
Explicitly, $\phi _n(y)$ can be written as
\begin{equation}\large
\phi _n(y)=\sin \pi ny \label{phifunc}
\end{equation}

Then for the expansion coefficients  we will have system of the
first order differential equations:
\begin{equation}\large
i\sum \dot {C}_n \phi_n=\sum C_n E_n \phi_n+\frac {1}{2}L^3 \ddot
{L}y^2\sum C_n \phi_n
 \label{eq2}
\end{equation}
\begin{figure}[t!]
\includegraphics[totalheight=0.23\textheight]{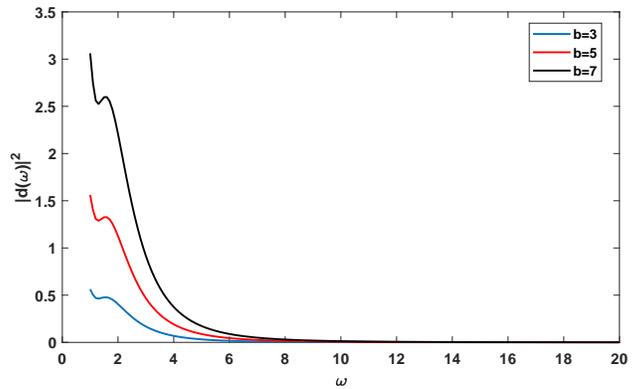}
 \caption{ (Color online) The Spectrum of particle in a time-dependent box as a function of frequency for different wall's initial positions for $\omega_0=1 $ and $a=10 $ } \label{fig:3}
\end{figure}

\begin{figure}[t!]
\includegraphics[totalheight=0.23\textheight]{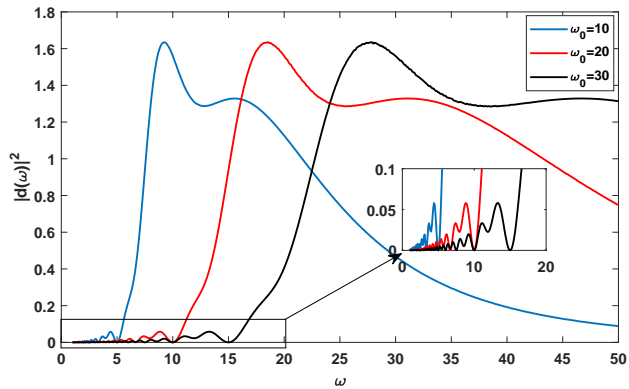}
 \caption{ (Color online) The Spectrum of particle in a time-dependent box as a function of frequency for different oscillation  frequencies for $a=10 $ and $b=5 $ } \label{fig:4}
\end{figure}

\section{High harmonic generation in time-dependent quantum box}

The dynamical confinement in the system, considered in the
previous section can be realized using optical field as the
confinement source. Typical versions of such systems are atom
optic billiards and optical tweezers. In atom optics billiards
rapidly scanning and tightly focused laser beam creating a
time-averaged quasi-static potential for confined particles
\ci{Raizen}-\ci{Montangero}. By controlling the deflection angles
of the laser beam, one can create various box (billiard) shapes.
In optical tweezers the radiation pressure from a highly focused
laser beam is able to create a dynamical trap for atoms and
subatomic scale  particles \ci{OTW1,OTW2}. In both cases
confinement boundary can be harmonically varying in time. The
interaction of the (confining) optical field with the particles of
a domain may induce different nonlinear optical phenomena,
including harmonic generation.  In the following we will consider
harmonically breathing box when $L(t)$ is given as
\begin{equation}\large
L (t)=a+b \cos\omega_0 t
 \label{interv}
\end{equation}

For the virtual system described by Eq.\re{shred03} the last term
can be written in the form of multichromatic field including 4
different freqencies as

$$
V(y,t)= \frac{1}{2}L^3 \ddot {L}y^2=$$

\begin{equation}
\frac{1}{2}(A+B\cos\omega_0 t+C\cos2\omega_0 t +D\cos3\omega_0
t+E\cos4\omega_0 t)y^2
\end{equation}
where
$$ A=-\frac{3b^2\omega_0^2}{8}(4a^2+b^2), \hspace{2mm}
B=-\frac{3ab\omega_0^2}{4}(4a^2+9b^2),$$

$$\hspace{2mm} C=-\frac{b^2\omega_0^2}{2}(3a^2+b^2), \hspace{2mm}
D=-\frac{3ab^3\omega_0^2}{4},  \hspace{2mm}
E=-\frac{b^4\omega_0^2}{8}$$

Thus the virtual system can be considered as a quantum box with
unit size and driven by nonlinearly polarized multichromatc field.
The average dipole moment for such system can be written as
\begin{equation}\large
\bar {d}= -\int_{0}^{L(t)} \psi^{*}(x,t)x\psi(x,t)dx = -2L\sum
_{m,n} C_{m}^{*}C_{n}V_{mn},\label{spektr}
\end{equation}

where
\begin{equation}\large
V_{mn}=\int_{0}^{1}\phi_{m}^{*}\,y\,\phi_n   dy
 \label{vvv}
\end{equation}

In Fig.3 $|d(\omega)|^2$ is plotted at different values of wall's
oscillation amplitude, $b$ and for the values of initial position,
oscillation frequency $a= 10$ and $\omega_0=1$, respectively.
Decay of the intensity when the harmonic order increases can be
seen from this plot, although for higher values of $b$ the decay
becomes slower.

Fig.4 presents plots of $|d(\omega)|^2$ at different values of the
wall's oscillation  frequency, $\omega_0$  for the values of the
wall's initial position and oscillation amplitude $a=10$ and $b=
5$, respectively. In general, the intensity decays rather slowly
(compared to Figs. 1-3), when the harmonic order grows. However,
the decay becomes very slow, for the higher values of $\omega_0$.
This result makes very attractive the dynamical traps, such as
quantum box with harmonically oscillating wall, from he viewpoint
of ultrashort pulse generation using optical high harmonic
generation.

\section{Conclusions}

In this paper we considered the problem of optical high harmonic
generation in confined quantum systems by considering the cases of
static and dynamic confinements. The static system represents 1D
quantum box driven by external linearly polarized monochromatic
field. The intensity of harmonic generation is analyzed by
computing numerically the average dipole moment as a function of
harmonic order. The analysis of the case of static confinement
shows that confinement does not lead to slowing done,  or growth
of the harmonic generation intensity.  In the case of dynamic
confinement the wall of the box is considered as oscillating and
created by an optical field, e.g. in atom optic billiards, or
optical tweezers, where the oscillating wall plays the role of
external field. It is found that when the wall's oscillation
frequency is large, the intensity of harmonic generation decays
very slow as a function of harmonic order. The model studied in
this paper can be useful for optical high harmonic and ultrashort
pulse generation under dynamical optical confinement using , e.g.,
atom optic billiards.

\end{document}